\input epsf
\documentstyle{mn}
\newif\ifAMStwofonts

\def\lesssim{\mathrel{\hbox{\rlap{\hbox{\lower4pt\hbox{$\sim$}}}\hbox{$<$}}}}
\def\gtrsim{\mathrel{\hbox{\rlap{\hbox{\lower4pt\hbox{$\sim$}}}\hbox{$>$}}}}
\def\apj{ApJ}

\def\aj{AJ}
\def\aap{A\&\hskip-1pt A}

\def\mnras{MNRAS}

\title[Identification of Bright Lenses]
      {Identification of Bright Lenses from the Astrometric \\
       Observations of Gravitational Microlensing Events}
\author[C.\ Han \& Y.\ Jeong]
       {Cheongho Han \& Youngjin Jeong \\
        Department of Astronomy \& Space Science, Chungbuk National University, 
	Chongju, Korea 361-763\\
	E-mail: cheongho,jeongyj@astronomy.chungbuk.ac.kr}
\date{Accepted
      Received }

\pagerange{\pageref{firstpage}--\pageref{lastpage}}

\begin{document}

\maketitle

\label{firstpage}

\begin{abstract}
When a source star is gravitationally microlensed by a dark lens,
the centroid of the source star image is displaced relative to the 
position of the unlensed source star with an elliptical trajectory.
Recently, routine astrometric follow-up measurements of these source 
star image centroid shifts by using high precision interferometers are 
proposed to measure the lens proper motion which can resolve the lens 
parameter degeneracy in the photometrically determined Einstein time 
scale.  When an event is caused by a bright lens, on the other hand, 
the astrometric shift is affected by the light from the lens, but one 
cannot identify the existence of the bright lens from the observed trajectory 
because the resulting trajectory of the bright lens event is also an 
ellipse.  As results, lensing parameters determined from the trajectory 
differ from those of a dark lens event, causing wrong identification 
of lens population.  In this paper, we show that although the shape 
and size of the astrometric centroid shift trajectory are changed due 
to the bright lens, the angular speed of centroid shifts around the 
apparent position of the unlensed source star is not affected by the 
lens brightness.  Then, one can identify the existence of the bright 
lens and determine its brightness by comparing the lens parameters 
determined from the `angular speed curve'  with those determined from 
the trajectory of observed centroid shifts.  Once the lens 
brightness is determined, one can correct for the lens proper motion.  
Since the proposed method provides both information about the lens 
brightness (dark or bright) and the corrected values of the physical 
parameters of the lens, one can significantly better constrain the 
nature of MACHOs.
\end{abstract}


\section{Introduction}
Searches for Massive Astronomical Compact Objects (MACHOs) by detecting 
light variations of stars caused by gravitational microlensing have been 
performed by several groups (MACHO: Alcock et al.\ 1993, 1996, 1997a, 
1997b; EROS: Aubourg et al.\ 1993, Ansari et al.\ 1996, Renault et al.\ 
1997; OGLE: Udalski et al.\ 1994, 1997; DUO: Alard \& Guibert 1997).  
With their efforts, more than 300 events have been detected toward the 
Galactic bulge and $\sim 20$ events toward the Magellanic Clouds.

Despite the large number of the detected events, the interpretations of 
the results of microlensing searches toward both fields are arguable.  
Toward the Galactic bulge, the determined microlensing optical depth of 
$\tau_{\rm GB} =3.9^{+1.8}_{-1.2}\times10^{-6}$ (Alcock et al.\ 1997a) 
substantially exceeds the optical depth of $\tau\sim 6\times 10^{-7}$ 
predicted for a standard Galactic disk (Paczy\'nski 1991; Griest et al.\ 
1991).  One explanation for the excess optical depth of bulge events would 
be the presence of undiscovered dark populations of disk lenses (Bahcall 
1984a, 1984b; Alcock et al.\ 1995).  Another possibility is that the 
lenses comprise an ordinary stellar population, and the excess optical 
depth is due to the existence of non-axisymmetric structure in the 
Galactic bulge (Kiraga \& Paczy\'nski 1994; Zhao, Spergel, \& Rich 1995).  
Toward the Magellanic Clouds, on the other hand, the combined microlensing 
optical depth of the MACHO and EROS survey of $\tau_{\rm LMC} 
= 2.1^{+1.3}_{-0.8}\times 10^{-7}$ (Alcock et al.\ 1997b; Renault et al.\ 
1997) is significantly below the expected optical depth of $\tau \sim 4.7
\times 10^{-7}$ for a halo composed entirely of MACHOs.  In addition, 
the time scales of the detected microlensing events indicate an average 
lens mass of $\sim 0.5\ M_\odot$.  The simplest interpretation of these 
results is that nearly half of the halo is composed of baryonic 
matter with a typical mass of order of $0.5\ M_\odot$ such as Hydrogen 
burning stars and white dwarfs.  However, both of these lens candidates 
have difficulties to reconcile with other observations (Gould, Bahcall, 
\& Flynn 1996; Gilmore \& Unavane 1999 for the stellar interpretation 
and Graff \& Freese 1996; Adams \& Laughlin 1996; Chabrier, Segretain, 
\& Mera 1996; Gibson \& Mould 1997 for the white dwarf interpretation), 
leading to alternative scenarios.  Sahu (1994) and Wu (1994) suggested 
that a large fraction of the Magellanic Cloud events could be due to 
microlensing by stars within the Magellanic Clouds themselves.  On the 
other hand, Zhao (1998a; 1998b) suggested that the Magellanic Cloud events 
may be due to lens objects in dwarf galaxies or tidal debris from a 
disrupted galaxy along the line of sight to the Magellanic Clouds.

The ambiguity in the nature of MACHOs arises because it is difficult to 
obtain information about the physical parameters of individual lenses
from the Einstein time scale, which is the only observable from current 
experiments.  Detection of microlensing events are accomplished through 
measurements of the variability of source star brightness caused by 
gravitational lensing effect.  The light curve of a lensing event is 
represented by
$$
A = { u^2+2\over u(u^2+4)^{1/2}};\qquad 
u = \left[ \beta^2 + \left( {t-t_0\over t_{\rm E}} \right)^2\right]^{1/2},
\eqno(1.1)
$$
where $u$ is the lens-source separation in units of the angular Einstein 
ring radius $\theta_{\rm E}$, and the lensing parameters $\beta$, $t_0$, 
and $t_{\rm E}$ represent the lens-source impact parameter, the time of 
maximum amplification, and the Einstein ring radius crossing time 
(Einstein time scale), respectively.  Once the light curve of an event 
is observed, these lensing parameters are obtained by fitting the observed 
light curve to the theoretical ones in equation (1.1).  One can obtain 
information about individual lenses because the Einstein time scale is 
related to the physical parameters of the lens by
$$
t_{\rm E} = {r_{\rm E}\over v};\qquad 
r_{\rm E} = \left( {4GM\over c^2}{ D_{ol}D_{ls} \over 
D_{os}}\right)^{1/2},
\eqno(1.2)
$$
where $r_{\rm E}=D_{ol}\theta_{\rm E}$ is the physical size of the 
Einstein ring radius, $v$ is the lens-source transverse speed, $M$ is 
the lens mass, and $D_{ol}$, $D_{ls}$, and $D_{os}$ are the separations 
between the observer, lens, and source star.  However, since the 
Einstein time scale depends on a combination of the lens parameters, 
the values of the lens parameters determined from it suffer from large 
uncertainties.  Therefore, to resolve the arguments in the interpretation 
of the result of the lensing experiments, a method that can resolve the 
lens parameter degeneracy for general microlensing events is essential.

A large part of the arguments about the nature of gravitational lenses 
can also be resolved if one can identify whether the lens is bright or 
dark.  If a large fraction of Galactic bulge events turn out to be caused 
by dark lenses, the most probable scenario for the observed Galactic bulge 
events will be that a significant fraction of Galactic matter in the disk 
is composed of dark lens population.  If similar results are obtained for 
Magellanic Cloud events, one can exclude the scenario of the self-lensing 
by stars in the Magellanic Clouds themselves.  Ideally, one can identify 
the existence of a bright lens because blended light from the lens will 
distort the shape of the microlensing event and shift the color of the 
observed star during the event (Buchalter, Kamionkowski, \& Rich 1996).  
In practice, however, it is very difficult to detect the presence of a 
blend by purely photometric means (Wo\'zniak \& Paczy\'nski 1997).  
Furthermore, since distortion and color change in the light curve can 
also occur due to other types of blend such as nearby unresolved blended 
stars and binary components (Han \& Kim 1999), detection of these symptoms 
of blending does not always imply the detection of a bright lens.  As a 
result, no event has been identified as a bright lens event by using 
this method.

Recently, routine astrometric follow-up observations of microlensing 
events with high precision interferometers are proposed as a method to 
partially break the lens parameter degeneracy of general microlensing 
events.  When a source star is gravitationally microlensed by a dark 
lens, the location of the apparent source star image centroid is displaced 
with respect to the position of the unlensed source star and the trajectory 
of the centroid traces out an ellipse (astrometric ellipse) during the
event (see \S\ 2).  Since the size of the astrometric ellipse, i.e.\  
the semi-major axis, is directly proportional to $\theta_{\rm E}$, 
one can determine the lens proper motion by $\mu=\theta_{\rm E}/t_{\rm E}$
combined with the photometrically determined Einstein time scale.
While the Einstein time scale depends on three lens parameters ($M$,
$D_{ol}$, and $v$),  the proper motion depends only on two parameters
($M$ and $D_{ol}$).  Therefore, by measuring the lens proper motion,
the uncertainty of the lens parameters can be significantly reduced
(Miyamoto \& Yoshii 1995; H\o\hskip-1pt g, Novikov \& Polnarev 1995; 
Walker 1995; Miralda-Escud\'e 1996; Paczy\'nski 1998; Boden, Shao, 
\& Van Buren 1998; Han \& Chang 1999).  When a source star is microlensed 
by a bright lens, on the other hand, the centroid displacement is 
distorted by the flux of the lens.  However, the resulting trajectory of 
the centroid shifts is also an ellipse, and thus one cannot identify the 
existence of the bright lens from the shape of the observed centroid 
shift trajectory (Jeong, Han, \& Park 1999).  In addition, the determined 
lensing parameters from the shape and size of the observed centroid shift 
trajectory will differ from the values for the dark lens case, causing 
wrong determination of lens parameters.

In this paper, we show that although the shape and size of the 
astrometric centroid shift trajectory is changed due to a bright lens, 
the angular speed of the source star image centroid shifts around the 
apparent position of the unlensed source star is not affected by the 
lens brightness.  Then, one can identify the existence of the bright 
lens and determine its brightness by comparing the lens parameters 
determined from the `angular speed curve' with those determined from the 
trajectory of observed centroid shifts.  Once the lens brightness 
is determined, one can correct for the lens proper motion.  Since the 
proposed method provides both information about the lens brightness 
(dark or bright) and the corrected values of the physical parameters of 
the lens, one can significantly better constrain the nature of MACHOs.

\section{Astrometric Shifts of Source Star Image Centroid}

\subsection{Trajectory of Astrometric Shifts}
When a source star is gravitationally microlensed, its image is split into 
two.  The typical separation between the two images for a Galactic event 
caused by a typical stellar mass lens is on the order of a milliarcsecond, 
which is too small for direct observation of individual images.  However, 
one can measure the astrometric displacements in the source star image 
light centroid by using several planned high-precision interferometers 
from space-based platform, e.g.\ the Space Interferometry Mission 
(http://sim.jpl.nasa.gov), and ground-based large telescopes, e.g.\ the 
Keck and the Very Large Telescope (Boden, Shao, \& Van Buren 1998).

For a dark lens event, the displacement vector of the source star image 
centroid with respect to the position of the unlensed source star is 
related to the lensing parameters by
$$
\vec{\delta\theta}_{c} = {\theta_{\rm E} \over u^2+2}
\left( {\cal T} \hat{\bf x} + 
\beta\hat{\bf y}\right); \qquad {\cal T} = {t-t_0\over t_{\rm E}},
\eqno(2.1.1)
$$
where $\hat{\bf x}$ and $\hat{\bf y}$ represent the unit vectors 
toward the directions which are parallel and normal to the lens-source 
transverse motion, respectively.  If we let $x=\delta\theta_{c,x}$ and 
$y=\delta\theta_{c,y}-b;\ b=\beta\theta_{\rm E}/2(\beta^2+2)$, the 
coordinates are related by
$$
x^2 + {y^2\over q^2} = a^2; 
\eqno(2.1.2)
$$
where 
$$
a = {\theta_{\rm E} \over 2(\beta^2+2)^{1/2}},
\eqno(2.1.3)
$$ 
and 
$$
q = {b\over a} = {\beta\over (\beta^2+2)^{1/2}}.
\eqno(2.1.4)
$$
Therefore, during the event the apparent source star image centroid traces 
out an ellipse (astrometric ellipse) with a semi-major axis $a$ and an 
axis ratio $q$.  Once the astrometric ellipse is measured, one can determine 
the lensing parameters $\beta$ and $\theta_{\rm E}$ because the axis ratio 
is related to the impact parameter and the semi-major axis is directly 
proportional to the angular Einstein ring radius.

\begin{figure}
\epsfysize=9.5cm
\centerline{\epsfbox{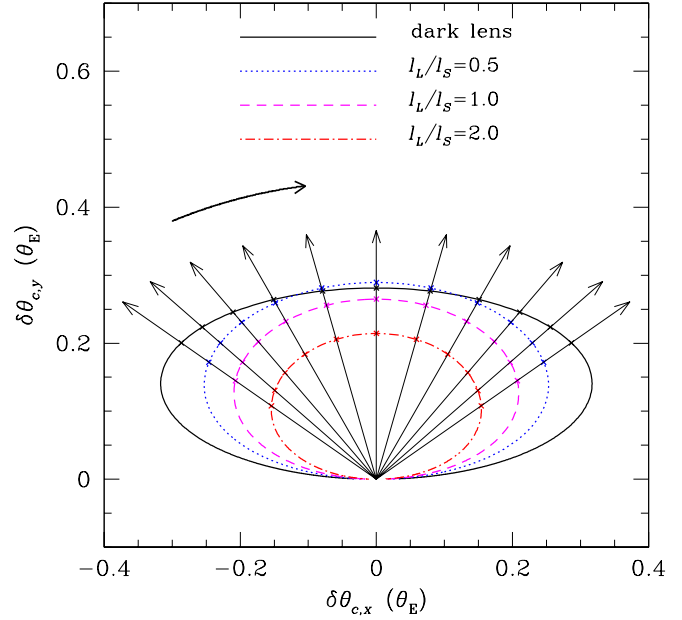}}
\caption{
Trajectories of source star image centroid shifts of gravitational 
microlensing events caused by bright lenses with various flux ratios 
between the lens and unlensed source star, $\ell_{L}/\ell_{S}$.  Each 
trajectory is for an example event with an impact parameter 
$\beta=0.7$.  We mark the position of the source star image centroid 
(`x') at several moments during the time interval $-1.0 t_{\rm E}\leq 
t\leq 1.0 t_{\rm E}$.  The arrow extending (with a thin solid line) 
from the origin represents the orientation vector of the centroid at 
each moment with respect to the apparent position of the unlensed source 
star at the origin.  The direction of centroid shift motion is marked 
by an arrow (with a thick solid line).
}
\end{figure}

If an event is caused by a bright lens, however, the centroid displacement 
is affected by the light from the lens, causing the observed centroid shifts 
differ from those for a dark lens event.  The bright lens affects the 
observed centroid shifts in two ways.  First, due to the flux of the lens, 
the apparent source star image centroid is shifted additionally toward the 
lens.  In addition, the reference position for the astrometric centroid shift 
measurement is not the position of the unlensed source star but the center 
of light between the unlensed source star and the bright lens.  By 
considering these two effects of the bright lens, the resulting centroid 
shifts of a bright lens event becomes
$$
\vec{\delta\theta}_c = {\cal D}
{\theta_{\rm E} \over u^2+2}({\cal T}\hat{\bf x} +
\beta\hat{\bf y}),
\eqno(2.1.5)
$$
where the deformation factor is related to the flux ratio between the lens 
and the unlensed source star $\ell_L/\ell_S$ by
$$
{\cal D} =
{
1+(\ell_L/\ell_S)+(\ell_L/\ell_S)\left[ (u^2+2)-u(u^2+4)^{1/2}\right]
\over
\left[ 1+(\ell_L/\ell_S)\right]\left[ 1+(\ell_L/\ell_S)
u(u^2+4)^{1/2}/(u^2+2)\right].
}
\eqno(2.1.6)
$$

In Figure 1, we present some example trajectories of source star image 
centroid shifts caused by bright lenses with various lens/source flux ratios
and they are compared to the trajectory for a dark lens event.  From the 
figure, one finds that the trajectory for the bright lens event is also 
an ellipse (observed astrometric ellipse).  Since both dark and bright 
lens events result in the same elliptical trajectories, one cannot 
identify the existence of a bright lens just from the shape of the observed 
astrometric ellipse.  One also finds that as the lens/source flux ratio 
increases, the shape of the observed astrometric ellipse becomes rounder 
and the size (measured by the size of the semi-major axis) becomes smaller.
Since the lensing parameters $\beta$ and $\theta_{\rm E}$ are determined 
from the shape and size of the observed astrometric ellipse, the determined 
lensing parameters for a bright lens event will differ from the parameters 
for the dark lens event (Jeong et al.\ 1999).

\begin{figure}
\epsfysize=11.5cm
\centerline{\epsfbox{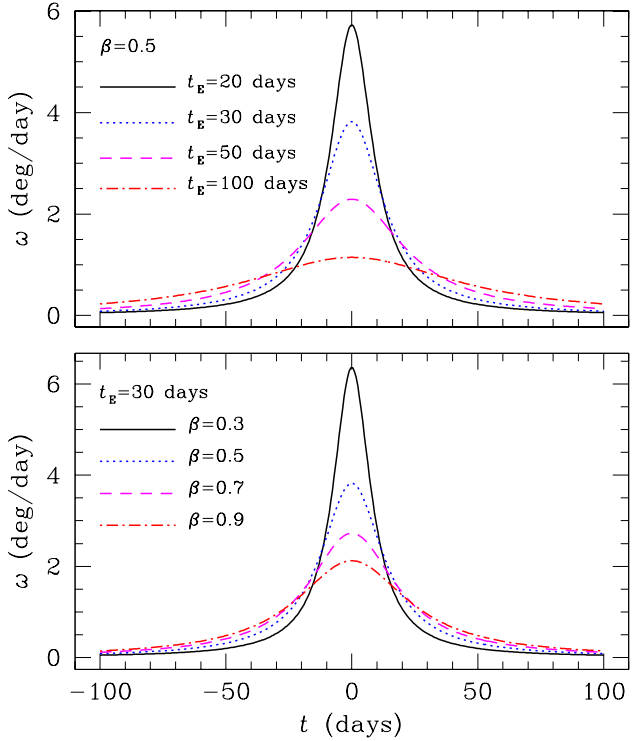}}
\caption{
Angular speed curves of the source star image centroid shifts 
for events with various values of the Einstein time scale $t_{\rm E}$
and the impact parameter $\beta$.  For all events, we adopt $t_0 = 0$.
}
\end{figure}

\subsection{Angular Speed Curve}
The motion of the source star image centroid is additionally characterized 
by its angular speed, $\omega$, around the apparent position of the unlensed 
source star.  The angular speed of the centroid for a dark lens event is 
determined by
$$
\omega(t) = {\partial\phi\over\partial t} = 
{\partial\phi\over\partial {\cal T}} 
{\partial {\cal T}\over\partial t} = 
{\beta t_{\rm E}\over (t-t_0)^2 + \beta^2 t_{\rm E}^{2}},
\eqno(2.2.1)
$$
where $\phi = {\rm Tan}^{-1}(\beta/{\cal T})$ is the orientation angle of 
the source star image centroid around the position of the unlensed source 
star.  In Figure 2, we present the changes in $\omega$ as a function of 
time (the angular speed curve) for events with various values of the 
Einstein time scale and the impact parameter.  From the figure, one finds 
that the angular speed increase as the source star approaches the lens 
and peaks when $t=t_0$.  In addition, the peak angular speed increases 
with decreasing Einstein time scale and impact parameter.  Since the 
angular speed is uniquely characterized by $\beta$ and $t_{\rm E}$, these 
lensing parameters can be determined from the measured angular speed curve.  
We note that among these lensing parameters the impact parameter can also 
be determined from the shape of the astrometric ellipse (see equation 
[2.1.4]).

\begin{figure}
\epsfysize=10cm
\centerline{\epsfbox{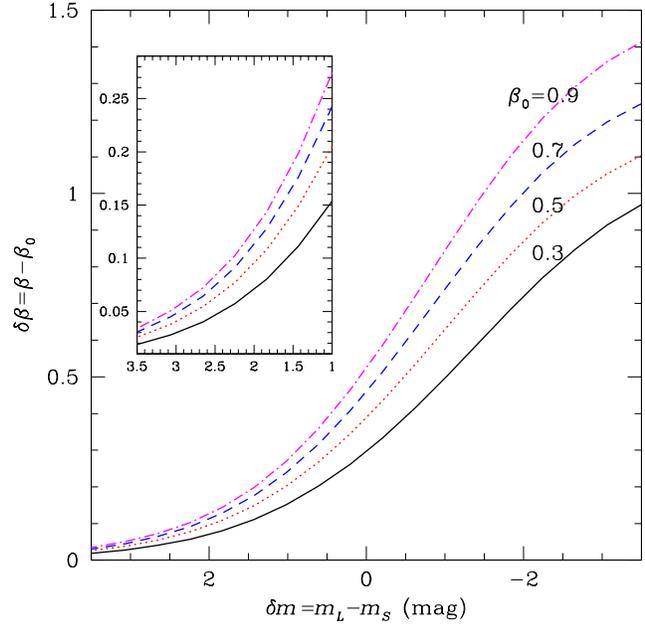}}
\caption{
Changes in the difference between the impact parameters determined from 
the observed astrometric ellipse, $\beta$, and the angular speed curve, 
$\beta_0$, as a function of the magnitude difference between the lens 
and source star, $\delta m$.  To better show  $\delta\beta=\beta-\beta_0$
in the region of small $\delta m$, we expand the region and present in a 
separate box.
}
\end{figure}

Another interesting finding about the motion of source star image centroid 
is that its angular speed is not affected by the bright lens.  When a 
source star is microlensed, the source star images with respect to the 
position of the unlensed source star lies at the same azimuth as the lens, 
i.e.\ lens-images are aligned along a straight line.  As a result, although 
the source star image centroid will experience extra amount of shift toward 
the bright lens along the straight line connecting the lens and the source, 
the orientation angle $\phi$, and thus the angular speed curve, will not 
be affected by the bright lens.  Since the angular speed curve for the 
bright lens event is same as that of the dark lens event, the lensing 
parameters for the bright lens event determined from the angular speed 
curve, $\beta_0$ and $t_{\rm E}$, will be the same as those for the dark 
lens event.  In Figure 1, we mark the positions of the source star image 
centroid (`x') at several moments during the time interval 
$-1.0t_{\rm E}\leq t\leq 1.0t_{\rm E}$ for events caused by bright lenses 
with various lens/source flux ratios.  We also mark the orientation vector 
of the centroid (an arrow with a thin solid line extending from the apparent 
position of the unlensed source star at the origin) at each moment.  One 
finds that regardless of the lens brightness, the orientation vectors at 
a same moment coincide.

Since the angular speed curve is not affected by the bright lens, while 
the trajectory of centroid shifts is affected by the lens, the value of 
the impact parameter of a bright lens event determined from the angular 
speed curve, $\beta_0$, will differ from the value determined from the 
observed astrometric ellipse, $\beta$.  Therefore, by comparing the impact 
parameters determined in two different ways, one can identify the existence 
of the bright lens and determine its flux.  Once the lens/source flux ratio 
is determined, one can also correct for the value of the lens proper 
motion.  In Figure 3, we present the changes in the impact parameter 
difference $\delta\beta=\beta-\beta_0$ as a function of the magnitude 
difference between the lens and source star, $\delta m = m_{L}-m_{S}$.  
To better show  the difference $\delta\beta$ in the region of small 
$\delta m$, we expand the region and present in a separate box.  From 
the figure, one finds that although $\delta\beta$ have different dependencies
on $\delta m$ for different values of $\beta_0$, the value of $\delta\beta$ 
is substantial even for events affected by lenses with low brightnesses.  
For example, the expected impact parameter difference for a bright lens 
event caused by a lens 2 mag fainter than the source star is 
$\delta\beta\sim 0.1$, which is well beyond the uncertainty of the 
astrometrically determined impact parameter (Boden et al.\ 1998).

\section{Summary}
We study the motion of the astrometric shifts of the source star image 
centroid caused by gravitational microlensing.  Findings from the study 
and the applications of the findings are summarized as follows.

\begin{enumerate}
\item
The angular speed of the source star image centroid increases as the 
source approaches the lens and peaks when $t=t_0$.  In addition, the 
peak angular speed increases with decreasing Einstein time scale and 
impact parameter.  Since the angular speed of an event is uniquely 
characterized by the lensing parameters $\beta$ and $t_{\rm E}$, these 
parameters can be determined from the measured angular speed curve.
\item
While the observed trajectory of the source star image centroid shifts 
is changed by the bright lens, the angular speed curve is not affected 
by the lens.  Therefore, although the lensing parameters determined 
from the observed centroid shift trajectory, i.e.\ $\beta$ and 
$\theta_{\rm E}$, differ from those for a dark lens event, the lensing 
parameters $\beta_0$ and $t_{\rm E}$ determined from the angular speed 
curve are same as those of the dark lens event.
\item
One can identify the existence of the bright lens and determine its 
brightness by comparing the impact parameters determined from the 
astrometric shift trajectory and the angular speed curve of the 
source star image centroid.  The difference in the two impact parameters 
is substantial even for events affected by lenses with low brightnesses, 
enabling one to easily identify the lens.
\end{enumerate}


\end{document}